\documentstyle[12pt,worldsci]{article}
\begin{document}

\begin{flushright}
OCIP/C 94-5 \\
hep-ph/9406278 \\
June 1994
\end{flushright}

\title{{\bf ISSUES IN LIGHT MESON SPECTROSCOPY: \\
THE CASE FOR MESON SPECTROSCOPY AT CEBAF}}
\author{Stephen Godfrey \\
{\em Ottawa-Carleton Institute for Physics \\
Department of Physics, Carleton University, Ottawa CANADA K1S 5B6 }}

\maketitle
\setlength{\baselineskip}{2.6ex}

\begin{center}
\parbox{13.0cm}
{\begin{center} ABSTRACT \end{center}
{\small \hspace*{0.3cm} I review some outstanding issues in meson spectroscopy.
 The most important
qualitative issue is whether hadrons with explicit gluonic degrees of
freedom exist.  To answer this question requires a much better understanding
of conventional $q\bar{q}$ mesons.   I therefore begin by
examining the status of conventional meson spectroscopy and how the
situation can be improved.  The expected  properties of
gluonic excitations are discussed with particular emphasis on
hybrids to give guidance to experimental searches.
Multiquark systems are commented upon as they are likely to be important
in the mass region under study and will have to be understood better.
In the final section I discuss the  opportunities that
CEBAF can offer for the study of meson spectroscopy.}}
\end{center}

\section{Introduction}

It is twenty years since the birth of Quantum Chromodynamics [1],
the theory of the strong interactions,
and it is not yet clear what the physical states of the theory are.  This
is an extraordinary statement that we still cannot answer such a basic
and fundamental question.
Although there is growing evidence for
the existence of hadrons with no valence quark content or with an excited
glue degree of freedom, we still do not really know if such states
exist.  An analogy in QED might be not knowing whether positronium
existed or not.  Clearly we cannot say that we understand QCD until these
questions are answered.
We have a powerful tool to better understand QCD through the
the interplay of theory and experiment.
A better understanding of QCD, a non-Abelian gauge
theory, will also lead to a better understanding of other
non-Abelian gauge theories such as the standard model of the electroweak
interaction.

Since it is possible that it may be further decades before we have a thorough
theoretical understanding of QCD  in the low $Q^2$ confinement region,
we must rely heavily on the insights
we can  gain from experiment and QCD based models.
To a large extent our understanding of hadron structure is based on the
constituent quark model
in which mesons are made of a quark and antiquark and baryons are made of
three quarks [2,3].
However, an important consequence of QCD is the expectation that {\it exotic}
hadrons beyond the naive quark model should also exist;
{\it hybrids, glueballs}, and $q\bar{q}q\bar{q}$ states.
Perhaps the discovery of such {\it exotica} will help us understand {\it Soft
QCD}. The problem is that although there are several exotic meson candidates,
no exotic has been unambiguously identified.  What has happened to them?
Answering this question has become the major preoccupation of hadron
spectroscopists.
In this contribution I review the expected properties of {\it conventional}
and {\it exotic} mesons and how these states might be studied at CEBAF.

As an operational definition I will refer to
states predicted by the constituent quark model as {\it conventional}
hadrons and those lying outside the quark model as {\it exotic} hadrons.
I emphasize, however,  that  there is
nothing fundamental about the quark model and the physical states of
the theory should be based on the gauge invariant operators one can
construct in QCD and will in general include gluonic excitations [4].
Nevertheless,  states
predicted by the quark model are the only ones that have been unambiguously
identified. Finding new types of hadronic matter --- the
{\it hybrids} which have constituent quarks and an excited
glue degree of freedom  and {\it glueballs} which have no valence quark
content what-so-ever [5] is the most important qualitative question in
hadron spectroscopy.
A serious impediment to the discovery of such states is
the sad shape of hadron spectroscopy.  In particular, none of the light
meson spectra is well mapped out for either orbital or radial excitations.
There are also numerous puzzles in
light meson spectroscopy, for example, the scalar meson puzzle, the nature
of the $f_1(1420)$ and the $f_0(1720)$, and the $g_T$ mesons.
The first priority is to sort out light meson spectroscopy so that we have
a template against which to compare exotic candidates.

Despite the shape of conventional meson spectroscopy
it is the discovery of gluonic excitations in the hadron spectrum which
is the most important issue in hadron spectroscopy as it will signify a
qualitative difference with the quark model.
The primary purpose of the next generation of hadron experiments
should be to discover glueballs, hybrids, and other exotic hadrons.
There are numerous models describing such states with important
qualitative differences so that
the discovery of exotics is important to distinguish between
the different models and make progress in understanding {\it soft} QCD.

\section{Conventional Mesons}

The Quark Model is 30 years old!  It is a useful tool for
understanding hadron spectroscopy but we still don't understand
why it works.  To make progress we need to fill in some of the missing
states so that we can either verify the model, find out where it needs to
be refined, or possibly show it is wrong.  In our quest for {\it exotic}
hadrons we should not forget that {\it conventional} $q\bar{q}$ mesons
can also tell us much about the nature of confinement.
For example, the linear Regge
trajectories of orbitally excited mesons is a consequence of the
linear confining potential [6] and the splittings of
orbitally excited multiplets reflects
the Lorentz structure of the confining potential [7].  The better we
understand hadrons, the more we can test the quark model and ultimately,
better understand QCD.

\subsection{Quark Model vs Experiment}

In the constituent quark model conventional mesons
are bound states of a spin $1\over 2$ quark and
a spin $1\over 2$ antiquark bound by a phenomenological
potential which  reflects the properties of QCD.
The quark and antiquark spins  combine to give total spin
$\vec{S}=\vec{S}_q + \vec{S}_{\bar{q}} = 0, \; 1 $
which is coupled to the orbital angular momentum $L$ to give states of
total angular momentum $\vec{J}=\vec{L}+\vec{S}$ resulting in
$J=L, \; L-1, \; L, \; L+1$.
This leads to meson parity and charge conjugation given by
\begin{equation}
P=(-1)^{L+1} \;\; {\rm and} \;\; C=(-1)^{L+S}
\end{equation}
resulting in the meson states of Table 1.
\begin{table}
\caption{The quantum numbers of the conventional $q\bar{q}$ mesons.}
\begin{center}
\begin{tabular}{|l|l|l|c|c|c|c|} \hline
&	& $J^{PC}$ & I=1   & I=0 ($n\bar{n}$) & I=0  $s\bar{s}$ & Strange \\ \hline
L=0 	& S=0 	& $0^{-+}$ & $\pi$ & $\eta$ & $\eta'$ & $K$ \\
	& S=1	& $1^{--}$ & $\rho$ & $\omega$ & $\phi$ & $K^*$ \\ \hline
L=1	& S=0	& $1^{+-}$ & $b_1$ & $h$ & $h'$ & $K_1$ \\
	& S=1	& $0^{++}$ & $a_0$ & $f_0$ & $f'_0 $ & $K_0^*$ \\
	&	& $1^{++}$ & $a_1$ & $f_1$ & $f'_1$ & $K_1$ \\
	&	& $2^{++}$ & $a_s$ & $f_2$ & $f'_2$ & $K_2^*$ \\  \hline
L=2	& S=0	& $2^{-+}$ & $\pi_2$ & $\eta_2$ & $\eta'_2$ & $K_2$ \\
	& S=1	& $1^{--}$ & $\rho_1$ & $\omega_1$ & $\phi_1$ & $K_1^*$ \\
	&	& $2^{--}$ & $\rho_2$ & $\omega_2$ & $\phi_2$ & $K_2$ \\
	&	& $3^{--}$ & $\rho_3$ & $\omega_3$ & $\phi_3$ & $K_3^*$ \\
	 \hline
. & . & . & . & . & . & . \\
. & . & . & . & . & . & . \\
. & . & . & . & . & . & . \\ \hline
\end{tabular}
\end{center}
\end{table}

A meson with $J^{PC}=1^{-+}$
would be forbidden in the constituent quark model and
since quarks have charge either $+2/3$ or $-1/3$ a doubly
charged meson, $m^{++}$, is also forbidden as a {\it conventional} meson
state.

Since we cannot at this point calculate hadron properties from first
principles we must rely on QCD motivated models to help interpret experimental
resonances.  Although there are many models in the literature the
constitutent quark model has the greatest success in describing hadron
properties.  In these models mesons are
described by a Schr\"odinger equation
\begin{equation}
H=T +V_{q\bar{q}}
\end{equation}
where $V_{q\bar{q}}$ is the effective quark-antiquark potential which
consists of a spin-independent confining potential and spin-dependent
terms.  The confining potential is typically of the form:
\begin{equation}
H^{conf} = -\frac{4}{3}\frac{\alpha_{s}(r)}{r} +br
\end{equation}
where the first term comes from one-gluon-exchange and the second term is
the linear confining potential.
The Coulomb piece dominates at short distance
becoming more and more important with increasing quark mass while the linear
piece dominates at large distance, becoming more important for the light
quark mesons.
This is illustrated in Fig. 1 where the rms radii of various mesons
are indicated on a plot of a QCD motivated potential.
One sees that mesons composed of the heavy $b$ quarks are
relatively small and sit in the Coulombic region of the potential, while
the lighter mesons sit in the linear region of the potential.  Thus,
the study of mesons constructed out of light quarks act as a probe of
confinement.
\begin{figure}
\vspace{7.0cm}
\caption{The $q\bar{q}$ potential.}
\end{figure}

The phenomenological spin dependent Hamiltonian is of the form:
\begin{equation}
H_{spin} = H^{hyp}_{ij} + H^{s.o.(cm)}_{ij} + H^{s.o.(tp)}_{ij}
\end{equation}
where
\begin{equation}
H^{hyp}_{ij}=\frac{4}{3}{{\alpha_s (r)}\over{m_i m_j}}
\left\{ { { {8\pi} \over{3} }
\vec{S}_i \cdot \vec{S}_j \, \delta^3 (\vec{r}_{ij})
+{1\over{r^3_{ij}} }
\left[ { { {3\vec{S}_i\cdot\vec{r}_{ij} \vec{S}_j \cdot\vec{r}_{ij} }
\over{r_{ij}^2}}-\vec{S}_{i} \cdot\vec{S}_{j} }\right] }\right\}
\end{equation}
is the colour hyperfine interaction,
\begin{equation}
H^{s.o.(cm)}_{ij}=\frac{4}{3}{{\alpha_s(r)}\over{r^3_{ij}}}
\left( { {1\over{m_i} } + {1\over{m_j} } }\right)
\left( { {{\vec{S}_i}\over{m_i} } +{{\vec{S}_j}\over{m_j}} }\right)
\cdot\vec{L}\quad
\end{equation}
is the spin-orbit
colour magnetic piece arising from the one-gluon exchange and
\begin{equation}
H^{s.o.(tp)}_{ij} =- {1\over{2r_{ij}} } \;
{ {\partial V(r) } \over {\partial r_{ij}} }\;
\left( { {{\vec{S}_i}\over{m_i^2}} + {\vec{S}_j\over
{m^2_j}} }\right)\cdot\vec{L}
\end{equation}
is the spin-orbit  Thomas precession term where $V(r)$ is the interquark
potential.
In these formulae $\alpha_s(r)$ is the running coupling constant of QCD.

The colour hyperfine interaction
is responsible for $^3S_1 - ^1S_0$ splitting in  $\rho -\pi$, $K^* -K$,
and $J/\Psi -\eta_c$.
The spin-orbit and tensor piece of the hyperfine interaction break the
degeneracy of the orbitally excited multiplets.
The spin-orbit interaction has two contributions,
$H^{s.o.(cm)}_{ij}$ and $H^{s.o.(tp)}_{ij}$.
Since the hyperfine term is
relatively short distance, it becomes less important for higher orbital
excitations.  Multiplet splittings then become a measure of the
spin-orbit splittings, with contributions of opposite sign coming from
the short range Lorentz vector one-gluon-exchange
and the long range Lorentz scalar linear confinement potential.
The ordering
of states within a multiplet of given orbital angular momentum gives
information on the relative importance of the two pieces.  For example,
for the L=1 strange meson multiplet, the quark model predicts
$M(^3P_2)> M(^3P_1) > M(^3P_0) $.
i.e. the $J=L+1$ member of the multiplet is more massive than the $J=L-1$
member.  In contrast, for the $L=4$ multiplet, the quark model predicts
$M(^3G_3)>M(^3G_4) >M(^3G_5)$
so now the ordering is inverted with the $J=L-1$ member more massive than
the $J=L+1$.  The reason for this is that higher L mesons have larger
radii so the linear part of the potential is more important than the short
distance Coulomb piece.  Although the details of this inversion are
model dependent, the inversion is a general property of QCD motivated
potential models. Thus the multiplet splittings act as a probe of the
confinement potential so that the study of
excited light quark system, such as ones with high orbital angular
momentum, provides information on non-perturbative QCD.

In what follows, for purposes of illustration,  I
will compare experimental data to the results of the particular model with
which I am most familiar and which constitutes a comprehensive
calculation of meson properties [8,9].
Let us start with the strange
meson spectrum which is shown if Fig. 2.  The spectrum is as rich as any
in atomic physics with a beautiful regularity and numerous transitions,
either electromagnetic transitions or strong transitions
via $\pi$ or $\rho$ emission.  There is
good agreement for the masses of the leading orbital excitations which
supports the picture of a linear confinement potential.  The
$^3S_1-^1S_0$ splitting is much smaller than the splitting between the
$^3P_J$ centre of gravity and the $^1P_1$ state which is consistent with
the expected properties of short distance one-gluon-exchange and a
Lorentz scalar confining potential.  A Lorentz vector confining potential
would lead to comparable splittings for $^3S_1-^1S_0$ and $^3P_J-^1P_1$.
The information decreases as we go to the higher orbitally excited
multiplets and for radial excitations.  A similar pattern is obtained
for the $s\bar{s}$ mesons except that they are even more sparsely mapped
out than the strange mesons.
\begin{figure}
\vspace{19.0cm}
\caption{The level diagram for strange mesons.  The wavy lines represent
$\gamma$ emission, the solid lines represent $\pi$ emission and the
dashed lines $\rho$ emission.}
\end{figure}

The status of of light meson spectroscopy is summarized in Fig. 3.
Starting with the P-wave multiplets, which is
the multiplet most filled,
we find that even it is not well understood.
The scalar mesons $(0^{++})$ are in a state of confusion due to the
possible interpretation of the $a_o (980)$ and $f_0 (980)$ as
$q\bar{q}q\bar{q}$ states[10-12].  In the $1^{++}$ sector the $f_1 (1420)$
has long been considered to be  the $^3P_1 \; s\bar{s}$ meson.
Recently the LASS group discovered another state [13] which appears a more
likely candidate, the $f_1(1530)$.
Finally, the $^1P_ 1 \; s\bar{s}$ state ($h'_1$)
is yet to be confirmed. Turning to the higher mass multiplets,
there is good agreement for the masses of the leading orbital excitations
between the quark model and experiment supporting the picture of linear
confinement.  In general, however, the radially excited and orbitally
excited mesons are even less understood than the P-waves.  In particular,
for the $s\bar{s}$ states there is very little known above the $L=1$
multiplet.
Some of my assignments  (or lack of assignments) in Fig. 3
are subject to debate but this lack of a consensus
underlines  the fact that far too little is known about light meson
spectroscopy.  Without completing at least some of these
multiplets we can hardly say we understand the meson spectrum.
\begin{figure}
\vspace{16.5cm}
\caption{Summary of mesonic states with light quarks.}
\end{figure}


\subsection{Some Puzzles in Mesons}

In addition to the obvious searches for the missing mesons there are
numerous puzzles in meson spectroscopy which may be hints of new
types of hadronic matter.  I will only mention some of them here and
refer the interested reader to other contributions and to the literature
for a more detailed account [14-16].

The $\eta(1440)$ (formerly the $\iota(1440)$)
is seen in the gluon rich $J/\psi$ radiative decay the $\iota$
and is therefore a prime candidate for a glueball [5].
It is now believed to be three separate states; two $0^{-+}$ and a
$1^{++}$ [15].  Two isoscalar
radially excited pseudoscalars ($2^1S_0$)
are expected to lie in this mass region so until we obtain a more complete
understanding of the $2^1S_0$ nonet the issue will remain cloudy.
Another
problem adding to the general confusion involves
the misidentification of the $E(1420)$.

The $f_0(1750)$ (formerly the $\theta(1750)$)  is also seen in
$J/\psi$ radiative decay to $K\bar{K}$ making it another
glueball candidate [5]. Although a number of $2^{++}$
states, both radially excited P-waves and orbitally excited F-waves
are expected, none seem to fit the $\theta$.  Further evidence
for its exotic character is the fact that it
was not seen in $K\bar{K}$ by the LASS group.  Dooley, Swanson and Barnes
speculate that the $\theta$ is a linear combination of loosely bound
$K^*\bar{K}^*$ and $\omega \phi$ pairs [17].

The ${f_1(1420)}$ (formerly the E(1420))
has for a long time thought to be the $1^3P_1$ $J^{PC}=1^{++}$
$s\bar{s}$ meson.
Recently the LASS group established the existence of another
axial vector meson at about 1530 MeV which appears to be a much stronger
candidate for this state [13].
If this is the case what is the E(1420)? and does this puzzle have anything
to do with the the $\iota (1440)$ puzzle?    Both states lie just above
$K^*\bar{K}$ threshold and perhaps we are again seeing some manifestation of
multiquark physics.

There are numerous puzzles besides the ones just mentioned.
For instance what is the explanation of
the $g_T$ $2^{++}$ tensor mesons seen at Brookhaven
in $\pi p \to K\bar{K}$ [18]?
Are they glueballs or are they conventional mesons
or do they have a totally different explanation?
Above, I mentioned that the $0^{++}$ $\delta(980)$ and $S^* (980)$ mesons
are thought to be $K\bar{K}$ molecules.
If this is indeed the case
where are the $^3P_0$ $q\bar{q}$ states?
Recently, candidates for the
$^3P_0$ $q\bar{q}$ states have been observed but they have yet to be
confirmed.

There are clearly many puzzles remaining in meson
spectroscopy which require both detailed experimental and theoretical
study.

\subsection{Hunting Missing States}

Given our unsatisfactory knowledge of light meson spectroscopy how do we
find the missing states and solve some of the puzzles?
One can see from Fig. 3 that as we go to higher
mass the number of states multiply rapidly so that in general we will
have to find the missing states in a large background of other
states.  It is therefore
highly unlikely that we will have any success in unravelling the
spectroscopy by bump hunting.  Rather, we will need high statistics
experiments to perform partial wave analysis to filter by $J^{PC}$
quantum numbers.  To assist us in this process a guide to the expected
properties will be useful and we refer to quark model predictions
for the expected masses and decay modes [8,9].
This can give us insight
into why some states are missing and how to look for them.
As an example of what such
a search would entail we examine some missing states
whose quark model predictions are listed in Tables 2 and 3.

\begin{table}
\caption{Quark Model predictions for the propeties of the missing L=2
mesons.  The masses and widths are given in MeV.
The masses come from Ref. 8 and the widths from Ref. 9.}
\begin{center}
\begin{tabular}{|l|l|l|} \hline
Meson State & Property  & Prediction \\ \hline
$\eta_2 (1^1D_2 )$ & Mass 			& 1680  	\\
		& width 			& $\sim 400$ \\
		& $BR(\eta_2 \to a_2\pi)$	& $\sim$ 70\%	\\
		& $BR(\eta_2\to \rho\rho)$	& $\sim$ 10\% \\
		& $BR(\eta_2\to K^*\bar{K}+c.c.)$	& $\sim$ 10\% \\
\hline
$\eta_2'\;(1^1D_2 )$ & Mass			& 1890 \\
		& width			& $\sim 150$ \\
		& $BR(\eta'_2 \to K^*\bar{K}+c.c.)$ & $\sim$ 100\% \\
\hline
$K_2 \; (1^1D_2 )$ & Mass 		& 1780  \\
		& width		& $\sim 300$ \\
		& $BR( K_2 \to K^* f(1280))$ & $\sim$ 30\%  \\
		& $BR( K_2 \to \rho K)$	& $\sim$ 20\%  \\
\hline
$\omega_1 \; (1^3D_1 )$ & Mass	& 1660	 \\
		& width		& $\sim 600$  \\
		& $BR( \omega_1 \to B\pi )$ & $\sim$ 70 \%  \\
		& $BR( \omega_1 \to \rho \pi )$ & $\sim$ 15\%	\\
\hline
$K_2 \; (1^3D_2 )$ & Mass		& 1810 	\\
		& width		& $\sim 300$  \\
		& $BR( K_2 \to K^* (1420)\pi)$ & $\sim$ 50 \%	\\
		& $BR( K_2 \to K^* \pi)$	& $\sim$ 30 \%	\\
\hline
$\rho_2 \; (1^3D_2)$ & Mass	& 1700  \\
		& width		& $\sim$ 500	\\
		& $BR( \rho_2 \to [a_2\pi]_S)$ & 	$\sim$ 55\%  \\
		& $BR( \rho_2 \to \omega \pi )$ &  $\sim$ 12\%	\\
		& $BR( \rho_2 \to \rho \rho )$ &  $\sim$ 12\% \\
\hline
$\omega_2\; (1^3D_2)$ & Mass	& 1700	\\
		& width		& $\sim 250$  \\
		& $BR(\omega_2 \to \rho \pi)$ & $\sim$ 60 \%	\\
		& $BR(\omega_2 \to K^*\bar{K} )$ & $\sim$ 20 \%     \\
\hline
$\phi_2\; (1^3D_2)$ & Mass		& 1910			\\
		& width		& $\sim$ 250		\\
		& $BR (\phi_2 \to K^*\bar{K}+c.c.)$ & $\sim$ 55\% 	\\
		& $BR (\phi_2 \to \phi \eta )$ & $\sim$ 25\%		\\
\hline
\end{tabular}
\end{center}
\end{table}

Starting with the $\eta_2 (1^1D_2)$ we expect it to be almost degenerate in
mass with its non-strange isoscalar partner, the $\pi_2$
(formerly the $A_3 (1680)$).
{}From Table 2 we see that it is expected to
be rather broad and it decays predominantly through the $a_2$ isobar which in
turn decays to $\rho\pi$.  The final state is expected to have
$4\pi$'s making  it  rather complicated to reconstruct
the original $\eta_2$ resonance.  The $\rho_2 (1^3D_2)$ will also  decay
dominantly to a $4\pi$ final state.  The $\omega_2$ decays to the simpler
$\rho\pi$ final state with a moderate width but since it has a similar
mass as the $\pi_2 (1680)$  which also decays
to $\rho\pi$ it is possible that it is masked by the $\pi_2$.  The
$s\bar{s}$ states, the $\eta_2' (1^1D_2)$ and the $\phi_2 (1^3D_2)$,
are both relatively narrow and one would expect that they would have been
observed.  In fact the LASS group has recently reported seeing
them in $K\bar{K}\pi$.
The likely reason that they have been so difficult to find is that they are
produced rather weakly.  The
strange mesons, the $K_2 (1^1D_2)$ and the $K_2 (1^3D_2)$ lie at
around 1800 MeV.  They decay to some relatively simple channels and
are predicted to have a moderate width.  It is possible that they have
been sited as the $L(1770)$'s.  The final state in the table is the broad
$1^{--}$ $\omega_1$.  Structure has been seen in this mass region but
the experimental situation is likely confused due to the nearby
broad $2^3S_1$ $1^{--}$ state which is expected to lie at around 1450 MeV
and  overlaps and interferes with the $1^3D_1$ state.

One can perform a similar analysis of other multiplets.  In Table 3 I give
the expected properties of some orbitally and radially excited mesons.
As listed,
most of these states appear to be relatively narrow with dominant
branching ratios to simple final states.  There are candidates for
several of these states.  For example, an $f_2$ is observed with mass
1810 MeV, width $\sim 200$ MeV (with large uncertainties) and
with $BR(f_2\to \pi\pi) \sim 20\%$.  Although the observed mass is consistent
with the quark model predictions, the predicted width is at least a factor
of two small. This discrepancy might be due to additional decay modes not
considered in the analysis, which only includes two body modes, or it
may reflect the sensitivity of the decay widths to the details of the
meson wavefunction where a slight shift in the node of the $2P$
wavefunction or possibly mixing with nearby states, can have a large
effect on the decay width.  Similarly, the observed mass
of the $K_2^*(1980)$ agrees with the quark model prediction and the decay
properties are not inconsistent with the quark model ($\Gamma^{theory}\simeq
150 \; {\rm MeV} \; vs \; \Gamma^{expt}\simeq 240 \; {\rm MeV}$).
Therefore, the predictions of Table 3 should be taken  as a rough
guide to the expected properties of radially excited P-wave mesons and
until they are more fully tested against experiment they should not be
taken too literally.  In addition, there is a need for further
theoretical work, to try to understand more complicated decay modes,
and model dependent effects on the meson properties.

\begin{table}
\caption{Quark Model predictions for the properties of some
of the N=2 P-wave mesons. The masses and widths are given in MeV.
The masses come from ref. 8 and the widths from ref. 9.}
\begin{center}
\begin{tabular}{|l|l|l|} \hline
Meson State & Property  & Prediction  \\
\hline
$a_2 (2^3P_2 )$ & Mass 			& 1820  		  \\
		& width			& $\sim 140$ 		  \\
		& $BR(a_2 \to \rho\pi)$		& $\sim$ 70\%		\\
		& $BR(a_2\to \eta\pi)$		& $\sim$ 10\%		\\
		& $BR(a_2\to K\bar{K})$		& $\sim$ 10\%		\\
		& $BR(a_2\to \eta'\pi)$		& $\sim$ 10\%		\\
		& $BR(a_2\to K^*\bar{K})$	& $\sim$ 10\%		\\
$a_2 (1^3F_2 )$ & Mass			& 2050			\\
\hline
$f_2\;(2^3P_2 )$ & Mass			& 1820			\\
		& width			& $\sim 90$		\\
		& $BR(f_2 \to \pi\pi )$		& $\sim$ 50\%		\\
		& $BR(f_2 \to K\bar{K})$	& $\sim$ 20\%		\\
		& $BR(f_2 \to K^*\bar{K})$ 	& $\sim$ 15\%		\\
$f_2\;(1^3F_2 )$ & Mass			& 2050			\\
\hline
$f_2'\;(2^3P_2 )$ & Mass			& 2040			\\
		& width			& $\sim 110$		\\
		& $BR(f_2' \to K\bar{K})$	& $\sim$ 35\%		\\
		& $BR(f_2' \to \eta\eta)$	& $\sim$ 10\%		\\
		& $BR(f_2' \to \eta'\eta)$	& $\sim$ 10\%		\\
		& $BR(f_2' \to K^*\bar{K})$ 	& $\sim$ 43\%		\\
$f_2'\;(1^3F_2 )$ & Mass			& 2240			\\
\hline
$K_2^* \; (2^3P_2 )$ & Mass 		& 1940 			\\
		& width			& $\sim$ 150		\\
		& $BR( K_2^* \to K \pi)$	& $\sim$ 20\%		\\
		& $BR( K_2^* \to K^*\pi)$	& $\sim$ 20\%		\\
		& $BR(K_2^* \to \rho K)$	& $\sim$ 20\%		\\
		& $BR(K_2^* \to K^*\pi)$	& $\sim$ 20\%		\\
$K_2^* \; (1^3F_2 )$ & Mass 		& 2150 			\\
\hline
\end{tabular}
\end{center}
\end{table}

\subsection{Future Directions}

{}From the preceeding  sections we conclude that we need a far better
understanding of meson spectroscopy before we can say that we
understand it and before we can exclude conventional interpretations of
an exotic candidate with conventional quantum numbers.
The first step is to find some of the missing states.
It will be important to fill in  both the orbitally excited
multiplets and the radially excited multiplets.  These missing states
will lie in a large background of other states.
To find them, results from the LASS spectrometer group show us that
we will need unprecedented statistics along with a partial wave analysis to
filter the $J^{PC}$ quantum numbers.

We will also need to develop new experimental and theoretical
techniques to study broad resonances. From the experimental side it
is clear that it will not be easy to identify a broad resonance in a
background of other broad resonances and in the presence of new
production thresholds.  From the theoretical side most
quark model calculations have treated mesons in the valence quark limit
without considering the influence of coupling to production and decay
channels.  We are at the point in our understanding that these effects
can no longer be ignored as they can make significant changes to the
observed hadron masses and decay properties. These effects are starting
to be examined by Barnes, Swanson, and Weinstein [19,20] .
In addition, we can no
longer ignore final state interactions. Some progress has been made on
this problem by Barnes and Swanson [21].
It will be
necessary to understand how to obtain observed cross sections starting
with the underlying spectrum when decay channel coupling is taken into
account along with final state interactions of the decay products.

\section{Gluonic Excitations [5]}

The complication in QCD which makes it so difficult to solve is the presence
of boson-boson interactions required by gauge invariance which in perturbation
theory gives rise to rather complicated three and four-boson couplings.
Although it is difficult to extract physical properties from the QCD
Lagrangian
it is these gluon self couplings which lead to the belief that gluons play
a dual role in QCD; as mediators of the strong force as in the
conventional $q\bar q$ mesons and $qqq$ baryons, and as constituents
in glueballs and hybrids. The problem at present is that it is not even
clear what the correct degrees of freedom are for soft QCD so
that the predictions of the different models must be viewed with caution.
Nevertheless, the discovery of glueballs or hybrids  would
be an important advance in our understanding QCD
in the ``soft'' region.

In searching for glueballs and hybrids there are two ways of
distinguishing them from conventional states:
\begin{enumerate}
\item To look for an excess of observed states over that predicted by the
quark model.  For this method to succeed
we need a very good understanding of conventional mesons
so that we can rule out a conventional interpretation of a newly found state.
The previous section has shown  the difficulties of
this approach given our incomplete knowledge of conventional mesons and
the general confusion in the 1.5 to 2.5 GeV mass region.
\item Search for exotic quantum numbers which would signal states that
cannot be conventional quark model states.
\end{enumerate}
Both approaches require a knowledge of expected glueball and hybrid
properties which we will examine in what follows.

We begin by sketching out two models of soft QCD
whose results we will use in what follows.  The qualitative differences
of these models stresses our ignorance of the low $Q^2$ regime of QCD.

\subsection{ The Bag Model [22-26]}

In the bag model [26]
hadrons are viewed as a region of space enclosing a fixed
number of quarks and gluons with
the model made Lorentz invariant by the
addition of a surface pressure term, $B_0$, to the Lagrangian density.
Inside the bag the quark fields, $\psi$, obey the free Dirac equation
along with the boundary conditions that; 1) there is no colour current through
the bag surface (S),
and 2) pressure balance determines the bag surface.
\begin{equation}
\begin{array}{rl}
(\not{p}-m)\psi =0 & \mbox{inside S} \\
\psi =0 & \mbox{outside S}
\end{array}
\end{equation}
The lowest energy
solutions have quarks in $1S_{1/2}$, $1P_{1/2}$, $1P_{3/2}$ eigenmodes.
Gluons in the Bag obey the free Helmholtz equation
subject to the same boundary conditions.
\begin{equation}
\begin{array}{rl}
(\nabla^2 +\omega^2)\vec{A}^a =0 & \mbox{inside S} \\
\vec{A}^a =0 & \mbox{outside S}
\end{array}
\end{equation}
The solutions are the transverse electric (TE)
and transverse magnetic (TM) cavity resonator modes
with $J^{PC}=1^{+-}$ and $J^{PC} = 1^{--}$ respectively.
In the zeroth order bag model, the mass of a hadron is simply the sum of the
quark and gluon constituent energies and the energy of the bag itself.
To go beyond the zeroth order bag model involves including contributions from
gluon exchange [22,23].

\subsection{The Flux Tube Model}

The flux tube model of hadrons  [27,28] is based on the strong coupling limit
of QCD [29]
with its parameters fixed from the familiar meson and baryon sectors.
The significant difference between the flux tube approach and the bag
model approach is that
the eigenstates of the strong coupling limit of (lattice) QCD consist of,
not quarks and gluons as in the bag model, but
quarks on lattice sites connected by arbitrary paths of flux links or in
the absence of quarks, of arbitrary closed loops of flux (glueloops).
It is assumed
that the flux tube picture survives departures from the strong coupling
limit or in other words, that the flux tubes do indeed form a reasonable set
of basis states, and that the adiabatic treatment of the flux tubes in the
presence of quark motion is reasonable.

In this picture the string states define adiabatic quark potentials
analogous to the nuclear potentials in  molecular
physics where adiabatic surfaces are defined for the nuclear motion based on
the faster moving electronic potentials.  We should then expect
a tower of quark states built on each string adiabatic surface.  This is
illustrated pictorially in Fig. 4.
In the flux tube model conventional hadrons correspond to gluonic fields
in the ground states.
\begin{figure}
\vspace{8.0cm}
\caption{The adiabatic quark potentials of the flux tube model.}
\end{figure}

\subsection{Pure Glue States}

Both the bag model and flux tube model expect that hadrons will exist with
no valence quark content at all.  The predictions for glueball masses vary
considerably from calculation to calculation.  I use the
flux tube model predictions as a guide because of their
agreement with lattice calculations [29,30].  These are shown in Fig. 5.  The
flux tube model predicts that the lowest glueloop (glueball)
is a $0^{++}$ at
1.5 GeV with all other states above 2 GeV and the lowest $J^{PC}$ exotic
at around 2.5 GeV.
For comparison, the bag model predictions are considerably lower
with $M_{0^{++},2^{++}}=1$ GeV, $M_{0^{++},2^{++}}=1.6$ GeV,
and $M_{0^{-+},2^{-+}}=1.3$ GeV.
Because of the uncertainty in the scalar meson sector it seems
likely that it would be very difficult to distinguish a scalar meson
from the poorly understood conventional mesons\footnote{Although the
excess of scalar mesons beyond the quark model predictions is seen as
evidence for glueballs in this sector}.
More generally, it will be difficult to
unambiguously determine that any state with conventional quantum
numbers is a glueball due to the dense background of conventional mesons.
The best bet will then be to find glueballs with exotic quantum numbers.
Unfortunately these states are all expected to have mass greater than 2.5
GeV and so will be difficult to find.  In addition, from the CEBAF point
of view, glueballs are not expected to couple strongly to either photons
or vector mesons so are likely to be difficult to produce at CEBAF.
Glueballs therefore do not seem to be
the best place to start our search for gluonic hadrons.
\begin{figure}
\vspace{10.0cm}
\caption{The low lying glueball mass spectrum. The flux tube
results come from ref. 27 and the lattice results from ref. 31.}
\end{figure}

\subsection{Hybrid Mesons}

{}From our previous discussion it appears that the most fruitful method
to search for hybrids is
to search for states with quantum numbers inconsistent with
quark model predictions.

The first step in this approach
is to enumerate the hybrid $J^{PC}$ quantum numbers.
To do this in a model independent manner  obeying gauge invariance
we form gauge invariant operators [32,33]
from a colour octet $q\bar q$ operator and a gluon field strength.
\begin{equation}
O = ( \bar{\psi} \Gamma \frac{\lambda_a}{2} \psi ) \otimes (\vec{E}^a \;
or \; \vec{B}^a )
\end{equation}
The resulting composite operator, known as an interpolating field,
is equally applicable to all approaches,
from the Bag-model to the flux tube model in addition to
more rigorous lattice gauge theory calculations.  For example the
interpolating field for the $1^{-+}$ state is given by
$(\bar{\psi}\vec{\gamma} \psi ) \times \vec{B}$ and for the $2^{+-}$ by
$(\bar{\psi}\vec{\gamma}\gamma_5 \psi ) \otimes \vec{B}$.
The quantum numbers of the low lying hybrids are given by:
$$\matrix{2^{++} & 2^{-+} & \underline{ 2^{+-}} & 2^{--} \cr
          1^{++} & \underline{1^{-+}} & 1^{+-} & 1^{--} \cr
	  0^{++} & 0^{-+} & \underline{0^{+-}} & \underline{0^{--}} \cr}$$
Higher J operators can also be constructed
but presumably they are higher in mass and more difficult to produce.
The underlined $q\bar{q}g$ states have exotic quantum numbers
not present in the constituent quark model.
If these exotic states are sufficiently low in mass
and do not have exceedingly large widths they could provide the {\it smoking
gun} evidence for hybrids which we are seeking:  Their discovery would
unambiguously signal hadron spectroscopy beyond the quark model.

\subsubsection{ Hybrid Masses}

\paragraph{ Bag Model Predictions:}
To make hybrids we combine a colour octet gluon
with a $q\bar{q}$ pair in a colour octet to obtain a colour singlet:
$$(q\bar{q})_8 \times g_8 = (q\bar{q}g)_1 + \ldots .$$
The lowest $q\bar{q}g$ hybrid meson multiplets are constructed from
a colour octet $q\bar{q}$ with $J^{PC} =0^{-+}$ or $1^{--}$, each in the
$J^{PC}=(1/2)^+$ mode, and a gluon in the lightest TE mode with $J^{PC}=1^{+-}$
resulting in the following lowest lying hybrids;
$$2^{-+},\; 1^{-+}, \; 1^{--}, \; 0^{-+}.$$
The SU(3) flavour quantum numbers of a hybrid are those of the component
$q\bar q$ pair so that hybrid mesons span the familiar SU(3) flavour nonets.
However, the I=0 and I=1 states are not degenerate because in the isoscalar
hybrids, the relative ease of internal annihilation of the $q\bar{q}$ pair
which is already in a colour octet, shifts the mass.

Some representative results of bag model calculations of the hybrid spectrum
which include spin-dependent forces due to gluon exchange [22-24]
are shown in Fig. 6 along with
constituent quark model predictions for conventional mesons.
Different calculations are in reasonable agreement
for the splittings but differ on the multiplet mass.
The $0^{-+}$, $2^{-+}$, and $1^{--}$ hybrid nonets are near in mass to
$q\bar{q}$ states with the same quantum numbers which can result in
considerable mixing.  This can only confuse the situation when determining
if a state is a hybrid or conventional meson.  Thus, the discovery of
such states would be difficult to be convincing because they are also
candidates for conventional states.

\begin{figure}
\vspace{17.0 cm}
\caption{Hybrid mass predictions.  The short dashed lines are the bag
model predictions of Barnes Close and deViron, ref. 22.  The shaded region are
the bag model predictions of Chanowitz and Sharpe for a range of values of
the quark and gluon self energies, ref. 23.  The long dashed lines are
the flux tube model predictons of Isgur and Paton, ref 27. The solid
lines are the conventional $q\bar{q}$ predictions of the relativized quark
model, ref. 8.}
\end{figure}

\paragraph{Flux Tube Model Predictions:}
There are two types of hybrids in this model, vibrational hybrids which
correspond to excitations of the quantum string into higher string
normal modes, and topological hybrids which have more complicated string
topologies and correspond to higher energy
adiabatic surfaces.  The latter are expected to be much higher in energy
so we will not discuss them further.

The adiabatic potentials are characterized by mode occupation numbers with
a polarization index and a string mode index.
The first excited state is doubly degenerate with
phonons of tranverse vibration with $\sigma=\pm 1$
angular momentum about the $q\bar{q}$ axis. When combined
with spin we get the 8 nearly degenerate nonets of hybrid mesons:
$$J^{PC} = 0^{\pm\mp} \quad 1^{\pm\mp} \quad 2^{\pm\mp} \quad 1^{\pm\pm}$$
with masses approximately $1.9\pm 0.1$ GeV for hybrids with no strange
quark content.  Among these states are
three $J^{PC}$ exotic nonets with nine neutral members having $J^{PC}=2^{+-}$,
$1^{-+}$, and $0^{+-}$.
These results should be contrasted to the bag model where there is no such
degeneracy because the TM mode is much higher in mass than the TE mode.
These results are compared to bag model results in Fig. 6.

\subsubsection{Hybrid Decays}

In the previous section we came to the conclusion
that the most promising approach for finding hybrids is to look
for ones with exotic quantum numbers.  Even so, there are numerous
states to consider so, for the sake of brevity,
we take the $\hat{\rho}$ and $\hat{\phi}\;1^{-+}$ exotics as examples.
Possible decays are given by:
\begin{eqnarray}
\hat{\rho} & \to & [\underline{\pi\eta},\underline{\pi\eta'},\pi\rho,
K^*\bar{K},\eta\rho,\ldots ]_P \nonumber \\
& \to & [\pi b_1, \pi f_1, \eta a_1, KK_1 \ldots ]_S \nonumber \\
\hat{\phi} &\to & [\underline{\eta\eta'}, k \bar{K}(1400),
	K^* \bar{K}, \ldots ]_P \nonumber \\
& \to & [\bar{K} Q_2]_S \nonumber \\
& \to & [\bar{K} Q_1]_D
\end{eqnarray}
The underlined
decays to two distinct pseudoscalars in a relative P-wave is a unique
signature of the $1^{-+}$ state.

Given this long list of decays we turn to the various models for guidance
to which modes are likely to be dominant.  One would naively expect S-wave
mesons in the $\pi\eta$ or $\pi\rho$ channels to be dominant
due to the large available phase space.  However, a common feature of
the various  models is the selection rule that the gluonic
excitation cannot usually transfer its angular momentum to the final
state meson pairs as relative angular momentum but must instead appear
as internal orbital angular momentum of the $q\bar{q}$
pairs.\footnote{I note however, that this selection rule
does not appear to be absolute.}
This eliminates $\pi\eta$, $\pi\rho$, $\pi\eta'$, $\eta\rho$, $\eta\eta'$,
and $K^*\bar{K}$.
The selection rule suppresses the decay channels which would likely be large
and may make hybrids stable enough to appear as conventional resonances
while at the same time explaining why hybrids with exotic $J^{PC}$
have yet to be seen;  they do not couple strongly to simple final states.

In particular,
in the Bag model the dominant decays occur when the valence gluon forms a
colour octet $J^{PC}=1^{+-}$ $q\bar{q}$ pair in which either $q$ or $\bar{q}$
is in a  P-wave mode.
The bag then contains two $q\bar{q}$ colour octets which after rearrangement
fall apart into $q\bar{q}$ singlets, one in an S-wave ground state with
$J^{PC}= 0^{-+}, 1^{--}$ and the other in an L=1 $J^{PC}=(0,1,2)^{++},1^{+-}$,
i.e. $J^{PC}=1^{-+} \to \pi f_1$ or $\eta a_1$.

The flux tube
model also predicts that the low lying hybrids will decay preferentially
to final states with one ground state S-wave and one excited P-wave
meson; $b_1(1235) \pi$, $a_2(1320) \pi$,
$K^*_2(1420) \bar{K}$, $\pi(1300) \pi, \ldots$, rather than two ground state
mesons like $\pi\pi$, $\rho\pi$, $K\bar{K}$.  The reason for this is
that the relative coordinates of the two final state mesons are parallel to
the initial meson axis and so cannot absorb the unit of string angular momentum
about the initial meson axis.  Hence the string angular momentum is
absorbed as an internal meson orbital angular momentum and the
selection rule is broken
for final states with different spatial wavefunctions [34].  The flux
tube model expects stronger coupling to final states with one S-wave and
one P-wave final state meson.

The flux tube model predictions are listed in
Table 4.  The flux tube model predicts that the $\hat{a}_2$,
$\hat{a}_0$, and $\hat{f'}_0$ are probably too broad to
appear as resonances.  The $\hat{\omega}_1$ decays mainly to
$[a_1 \pi]_S$ and $[\pi(1300) \pi ]_P$ with $\Gamma\sim 100$ MeV
which would make it difficult to
reconstruct the original hybrid given the broad widths of the final state
mesons.  Similar problems also make the
$\hat{\phi}_1$ difficult to find.  According to the flux tube model
the best bets for finding hybrids are:
$\hat{\rho}_1$, $\hat{f}_2$, $\hat{f}_0$, and $\hat{f'}_2$.

\begin{table}
\caption{
The dominant decays of the low-lying exotic hybrid mesons.  From ref. 28.}
\begin{center}
\begin{tabular}{|l|l|l|} \hline
Hybrid State & $[\hbox{decay mode}]_{\hbox {L of decay}}$
	& Partial Width (MeV) \\
\hline
$\hat{a}^{+-}_2 (1900)$ & $\to [\pi a_2]_P$    & 450 \\
			   & $\to [\pi a_1]_P$    & 100 \\
 			   & $\to [\pi h_1]_P$    & 150 \\
$\hat{f}^{+-}_2 (1900)$ & $\to [\pi b_1]_P$  & 500 \\
$\hat{f'}^{+-}_2 (2100)$ & $\to [K\bar{K}^*_2 +c.c.]_P$  & 250 \\
			   & $\to [\bar{K}Q_2 +c.c.]_P$  & 200 \\
$\hat{\omega}^{-+}_1 (1900)$ & $\to [\pi a_1]_{S,D}$  & 100,70 \\
			   & $\to [\pi \pi(1300)]_P$  & 100 \\
			   & $\to [\bar{K} Q_2 +c.c.]_S$  & 100 \\
$\hat{\rho}^{-+}_1 (1900)$ & $\to [\pi b_1]_{S,D}$  & 100, 30 \\
				& $\to [\pi f_1]_{S,D}$  & 30, 20 \\
$\hat{\phi}^{-+}_1 (2100)$ & $\to [\bar{K} Q_1  + c.c.]_D$  & 80 \\
			& $\to [\bar{K} Q_2  + c.c.]_S $ & 250 \\
			& $\to [\bar{K} K (1400) + c.c.]_P$  & 30 \\
$\hat{a}^{+-}_0 (1900)$   & $\to [\pi a_1]_P $ & 800 \\
			     & $\to [\pi h_1]_P$  & 100 \\
			     & $\to [\pi \pi (1300) ]_S$ & 900 \\
$\hat{f}^{+-}_0 (1900)$ & $\to [\pi b_1]_P$  & 250 \\
$\hat{f'}^{+-}_0 (2100)$   & $\to [\bar{K} Q_1 +c.c. ]_P$ & 800 \\
			     & $\to [\bar{K} Q_2 +c.c. ]_P$ & 50 \\
			     & $\to [\bar{K} K(1400) +c.c. ]_S$ & 800 \\
\hline
\end{tabular}
\end{center}
\end{table}

What we conclude from all this is that the favoured final states
all contain broad P-wave mesons.  To reconstruct the original resonance
an isobar analysis will be essential and to do this we will again need
unprecedented statistics to pull a signal from the background.

\section{Multiquark States}

Multiquarks are discussed in detail in the contribution of Weinstein [20].
Here I comment briefly on some points relevant to meson spectroscopy.
Upon considering $qq\bar{q}\bar{q}$ systems we find that
the colour couplings are not unique as they are in mesons and baryons
and whether or not  multiquark states exist is a dynamical question.  It is
possible that multiquark states exist as bound states [35]
but it is also possible that
$qq\bar{q}\bar{q}$ configurations lead to meson-meson potentials [12].  Both
must be taken into account when attempting to unravel the meson spectrum.

A study of the $J^{PC}$ sector of the $qq\bar{q}\bar{q}$ system found
that weakly bound $K\bar{K}$ ``molecules'' exist in the isospin zero and
one sectors in analogy to the deuteron.  It was suggested that these
two bound states be identified with the $f_0 (975)$ and $a_0 (980)$ (the
$S^*$ and $\delta$).
The meson-meson potentials which come from this picture,
when used with a coupled channel Schrodinger equation, reproduce the
observed phase shifts for the $\delta$ and $S^*$ in $\pi\pi$ scattering [12].
The $K\bar{K}$ molecules are the exception however, as the model predicts
that in general the $qq\bar{q}\bar{q}$ ground states are two unbound mesons.

There is evidence that meson-meson potentials must be considered in other
processes as well. In the reaction $\gamma\gamma\to \pi^+ \pi^-$
the meson-meson potentials are needed along with $q\bar{q}$
resonances to reproduce the  $\gamma\gamma \to \pi^+\pi^-$
cross section data [36].  Enhancements in the production of low
invariant mass $\pi\pi$ pairs have been observed in a number
of processes; $\eta' \to \eta \pi\pi$, $\psi' \to J/\psi \pi\pi$,
$\Upsilon(nS) \to \Upsilon (mS) \pi\pi$, and $\psi\to \omega\pi\pi$.
Similar enhancements have also been seen in some $K\pi$ channels
in $\bar{p}p\to K\bar{K}\pi$.  The conclusion drawn from
these examples is that final state interactions arising from meson-meson
potentials will play a central role in understanding the 1 to 3 GeV mass
region.
So far only  pseudoscalar mesons in the final state have been considered
so the next logical step is to extend the analysis to vector-vector
and pseudoscalar-vector channels.  Perhaps these multiquark effects
are the key to the $E/\iota$ and $\theta$ puzzles.

\section{Meson Spectroscopy at CEBAF}

CEBAF offers a number of possibilities for studying meson spectroscopy
using high intensity photon beams incident on nuclear targets.
Using the CEBAF electron beam
high energy photons can be produced by either bremstrahlung
through a thin radiator
or backscattering a high powered laser from the incident electron beam
(a ``Compton Collider'').  The resulting photon energy will be close to
the original beam energy resulting in center of mass energies
ranging from $\sim 4$~GeV for an 8~GeV electron beam to $\sim 4.8$~GeV
for a 12~GeV incident electron beam.
The photons can be used to photoproduce meson resonances covering the
poorly explored region of 2 to 4 GeV where many conventional
and non-conventional mesons are expected to lie.  In particular, the
lowest lying hybrid mesons which are expected at around 2 GeV.

The basic production mechanism is that, through vector meson dominance,
the photon has vector meson components such as the $\rho$, $\omega$, and
$\phi$ so that the nucleon target is interacting with the vector meson
component of the photon.  There are numerous ways that the vector meson
can then interact with the nucleon target to  produce excited final state
mesons
which differ primarily by the t-channel exchange mechanism.  These are
illustrated in Fig. 7. Excited states
can be produced via diffraction --- the exchange of a (colourless)
pomeron; inelastic
production where the original target nucleon is excited into, say, an
$N^*$; and charge exchange where a charged pion is exchanged in the t-channel
so that the target nucleon is excited into, for example, a $\Delta^{++}$.
In addition, the photon can interact as a photon through Primakof production
where,
the photon excites a t-channel $\pi$ to produce the final state meson.
Finally, the photon can interact with a t-channel photon to produce a
final state resonance via two photon fusion.   Taken together CEBAF
offers a wide range of complementary production mechanisms which can help
decipher the underlying meson structure.
\begin{figure}
\vspace{10.0 cm}
\caption{ Meson production mechanisms at CEBAF.}
\end{figure}

Because the
photons have a relatively large $s\bar{s}$ content they are a good source of
strangeonium states.  Because the $s$-quark is intermediate in mass
between the heavier $c$ and $b$ quarks where we believe that quark
potential models are reasonable approximations to QCD
and the lighter $u$ and $d$ quarks where relativistic effects make the
naive quark models suspect, strangeonium spectroscopy provides a useful
bridge between these two extremes.  Thus, CEBAF can add considerably to
our knowledge of both the radially excited and orbitally excited
strangeonium states which are important for our understanding of soft
QCD.

In addition, it has been speculated that photoproduction experiments are
a good place to search for hybrids [28].  The basic idea is that
the glue in the vector mesons is excited by the t-channel particle
exchange to produce a hybrid meson.  However, detailed calculations of
this production mechanism do not presently exist and are only now being
performed [37].

In summary, CEBAF offers some interesting production mechanisms for
mesons. With the extremely high intensity of the electron beam, and hence
of the photon beam it offers the possibility of the very high statistics
needed for the next generation of meson spectroscopy experiments.  It may
turn out that CEBAF will be the long awaited KAON factory.

\section{Final Comments}

I hope I have demonstrated that meson spectroscopy is an extremely rich
subject with fundamental unanswered questions.
Our present knowledge of hadron spectroscopy
is a very shaky foundation on which to base our understanding of QCD.
At present it is not even clear what the relevant
degrees of freedom are for describing this regime of QCD.
The first step to understanding {\it soft} QCD is to find the
missing conventional $q\bar{q}$ states.
Until we understand conventional hadrons better  it will be very difficult
to make progress in finding  evidence for the gluonic degree of freedom
in the hadron spectrum which is the outstanding issue.

Although it is conceivable that hybrid states with non-exotic quantum
numbers could be identified as being excess states beyond those predicted
by the quark model, given the very broad range of predictions for
hybrid masses, I very much
doubt that this is the most fruitful approach.  It is more likely that, to
be successful, a hybrid search should focus on the exotic properties of
hybrids which would offer unambiguous evidence of new physics.
The most uncontroversial  such characteristic is that all models
agree that one of the lowest hybrids will have
exotic $J^{PC}$ quantum numbers $1^{-+}$ with mass
about $1.6\pm 0.3$ GeV.  The presence of a resonance signal
in this channel would be strong evidence for the discovery
of a hybrid so it seems sensible that this be the place
to begin any experimental search.

One should appreciate that the study of
$q\bar{q}$, hybrids, glueballs, and $qq\bar{q}\bar{q}$
is an indivisable
subject since they are all governed by the same theory --- QCD, and
require the same experiments.
To unravel the meson spectrum in the 1 to 3 GeV  mass region
will take unprecedented statistics.
It is important that many hadron properties be studied
in many different channels.
CEBAF has an important role in meson physics.  Because of the high
$s\bar{s}$ content of the photon it will be able to produce large numbers
of $s\bar{s}$ states significantly improving our incomplete knowledge of
this sector.  CEBAF also offers the possibility of discovering hybrid
mesons which would provide evidence for gluonic excitations in mesons.

\section*{ACKNOWLEDGEMENTS}

I would like to thank the organizers of the Workshop on CEBAF at Higher
Energies for
the invitation to participate in a very lively and enjoyable meeting.
In particular,  I would like to thank Ted Barnes and Jim Napolitano for
their kind invitation to speak.  I am looking forward to many future
meetings on meson physics at CEBAF.
This work was supported  by the Natural Science and
Engineering Research Council of Canada.

\bibliographystyle{unsrt}

\begin{thebibliography}{99}
\bibitem{[34]}
The selection rule was first pointed out by the Orsay group in the context
of constituent gluon models.
\end{thebibliography}

\end{document}